# Indirect exciton-phonon dynamics in $MoS_2$ revealed by ultrafast electron diffraction

Jianbo Hu[1,2,4,*], Yang Xiang[1], Beatrice Matilde Ferrari[3], Emilio Scalise[3], and Giovanni Maria Vanacore[3,4,*]

1. Laboratory for Shock Wave and Detonation Physics, Institute of Fluid Physics, China Academy of Engineering Physics, Mianyang 621900, China
2. State Key Laboratory for Environment-Friendly Energy Materials, Southwest University of Science and Technology, Mianyang 621010, China
3. Department of Materials Science, University of Milano-Bicocca, Milano (Italy)
4. Physical Biology Center for Ultrafast Science and Technology, California Institute of Technology, Pasadena (CA), USA.

*To whom correspondence should be addressed: jianbo.hu@caep.cn, giovanni.vanacore@unimib.it

**Transition metal dichalcogenides layered nano-crystals are emerging as promising candidates for next-generation optoelectronic and quantum devices. In such systems, the interaction between excitonic states and atomic vibrations is crucial for many fundamental properties, such as carrier mobilities, quantum coherence loss, and heat dissipation. In particular, to fully exploit their valley-selective excitations, one has to understand the many-body exciton physics of zone-edge states. So far, theoretical and experimental studies have mainly focused on the exciton-phonon dynamics in high-energy direct excitons involving zone-center phonons. Here, we use ultrafast electron diffraction and ab initio calculations to investigate the many-body structural dynamics following nearly-resonant excitation of low-energy indirect excitons in $MoS_2$. By exploiting the large momentum carried by scattered electrons, we identify the excitation of in-plane K- and Q-phonon modes with $E'$ symmetry as key for the stabilization of indirect excitons generated via near-infrared light at 1.55 eV, and we shed light on the role of phonon anharmonicity and the ensuing structural evolution of the $MoS_2$ crystal lattice. Our results highlight the strong selectivity of phononic excitations directly associated with the specific indirect-exciton nature of the wavelength-dependent electronic transitions triggered in the system.**

In this paper, we report on the many-body structural dynamics of $MoS_2$ following nearly-resonant excitation of low-energy indirect excitons. By means of ultrafast electron diffraction and ab initio calculations, we demonstrate that in-plane K- and Q-phonon modes with E' symmetry are key for the stabilization of indirect excitons in $MoS_2$, and we elucidate the role played by the phonon anharmonicity in the ensuing structural evolution of the crystal lattice.

## INTRODUCTION

The understanding and active control of material properties are crucial for addressing the technological challenges of the 21st century, associated with the demands for data processing, storage, and transmission. In such a context, transition metal dichalcogenides (TMDs) are showing great promise to satisfy these demands due to their unique physical properties [1], resulting from a subtle balance between electronic, lattice, and valley degrees of freedom [2,3]. The ability to shift such balance would provide a promising way to manipulate their behavior, thus offering unprecedented opportunities for optoelectronics as well as quantum technologies.

In equilibrium conditions, the system mainly explores thermodynamically stable states described within the Ehrenfest's scheme. Instead, perturbing the free-energy landscape using femtoseconds (fs) electromagnetic fields can lead the system toward transient states that are not accessible under equilibrium ergodic conditions [4]. Controlling these states and dynamically engineering their appearance on a microscopic level would allow to manipulate the macroscopic functionality of the system. This is particularly relevant for many-body exciton physics that exploits the valley-selectivity typical of TMDs [5,6,7,8,9], with a profound technological impact on the design of next-generation gates, memories, photon emitters as well as thermoelectric devices.

In the non-equilibrium regime, the Born-Oppenheimer approximation fails to describe the state of the system, and interactions between electronic and atomic degrees of freedom have a crucial role in determining their fundamental properties, such as electron mobility [10], quantum coherence loss and dephasing [11,12], heat generation and dissipation [13]. Thus, a complete understanding is only obtained when capturing such electronic-driven structural dynamics at the proper length (atomic) and temporal (femtosecond) scales. Here, we adopted ultrafast electron diffraction (UED) to investigate the exciton-phonon dynamics in molybdenum disulfide ($MoS_2$), the prototypical TMD. Following nearly-resonant excitation of low-energy indirect excitons at the Q and K points of the Brillouin zone, we observed a structural evolution that is consistent with the phonon anharmonicity resulting from the excitation of in-plane Q- and K-phonon modes with $E'$ symmetry. Our observations expand previous studies of exciton-phonon dynamics in $MoS_2$ reporting on the strong coupling of direct excitons with out-of-plane $A_{1g}$ Γ-phonons [5,6]. The direct investigation of atomic displacements in real-time has become possible only recently thanks to the development of highly-sensitive ultrafast electron and X-ray diffraction techniques [14,15,16,17,18]. By exploiting the large momentum carried by diffracted electrons, UED allows to explore short-range and long-range atomic-scale motions of the lattice [19,20,21,22,23,24,25,26,27,28], as well as zone-edge atomic vibrations that are not detectable with the traditional optically-based techniques, which are instead only sensitive to zone-center phonons because of the small momentum of light.

## RESULTS AND DISCUSSION

Nanoflakes of single-crystals $MoS_2$ were obtained via mechanical exfoliation from a natural $MoS_2$ bulk crystal, and then transferred onto a TEM Cu grid. The schematic of the UED experiment is shown in Fig. 1a: ultrashort electron pulses are focused in normal incidence on the $MoS_2$. The dynamics is initiated by ultrashort laser pulses with a variable wavelength (either 800 nm or 400 nm), temporal duration of 120 fs, repetition rate of 2 kHz, and with a fluence of 11.8 $mJ/cm^2$ or 5.9 $mJ/cm^2$ for the two wavelengths, respectively. The diffracted electrons are then recorded on a CCD detector in the stroboscopic mode at different delay times between the excitation laser and the probing electron pulse. Further details on the sample preparation and experimental setup are given in the Supporting Information. It is worth noting that at the adopted excitation intensity the effect of two-photon absorption (TPA) for illumination with light at 800 nm can be considered negligible since its contribution is about one order of magnitude smaller than the one-photon absorption [29,30].

$MoS_2$ is probably the most studied compound of the transition metal dichalcogenide family. It is characterized by S-Mo-S single layers bonded together by weak van der Waals interactions. Each layer consists of two hexagonal planes of S atoms intercalated by one hexagonal plane of Mo atoms. In the 2H polytype, Molybdenum is bound with Sulfur via a trigonal prismatic arrangement (space group is P3m1 and point group is $D_{6h}$). Fig. 1b shows a representative diffraction pattern measured before laser excitation, showing the in-plane hexagonal pattern typical of a $D_{6h}$ crystal where we can identify three high-symmetry lattice plane families: $\{1\bar{1}00\}$, $\{01\bar{1}0\}$, and $\{10\bar{1}0\}$.

Bulk 2H-$MoS_2$ is an indirect semiconductor with a bandgap of 1.29 eV at the Q point along the Γ-K direction of the Brillouin zone (see Fig. 1c). The direct gaps in Γ and in K are, instead, above 2 eV. This means that optical excitation at 400 nm (3.1 eV) can induce strong vertical interband electronic transitions across almost the entire Brillouin zone (see blue arrows in Fig. 1c) with no momentum change required for the electron. Instead, in the case of optical excitation at 800 nm (1.55 eV), the only allowed transitions would be non-vertical interband electronic transitions mediated by the presence of a phonon necessary to fulfill the momentum conservation (see red arrows in Fig. 1c). Such wavelength-dependent physics thus gives us an unprecedented opportunity to dynamically investigate the coupling between electronic and phononic degrees of freedom in a 2D material.

A direct consequence of such an effect is the formation of indirect excitons following the 800 nm optical excitation. In multi-layers $MoS_2$, photoluminescence (PL) spectroscopy has evidenced the presence of a well-defined feature in the range 1.4 – 1.6 eV [31], which was associated with indirect radiative recombination processes. In the bulk 2H-$MoS_2$, momentum-resolved electron energy-loss spectroscopy measurements have observed a strong excitation dispersing from 1.46 eV to 1.57 eV along the Γ-K direction [32] for a momentum value of the order of $q_K$ = 1.33 $Å^{-1}$, which corresponds to the momentum-space difference between the Γ and the K points of the Brillouin zone. The authors thus attributed such feature to an indirect $\Gamma_V$-$K_C$ exciton (where the subfix V and C stand for valence and conduction, respectively). Moreover, ab initio calculations predicted the existence of indirect $\Gamma_V$-$K_C$ excitons in the range 1.33 eV – 1.59 eV and indirect $\Gamma_V$-$Q_C$ excitons in the range 1.23 eV – 1.44 eV [33,34].

In our experimental configuration, we used near-infrared laser light at 800 nm (1.55 eV) to induce a nearly-resonant excitation of such indirect $\Gamma_V$-$K_C$ and $\Gamma_V$-$Q_C$ excitons, while light at 400 nm (3.1 eV) would mainly promote the formation of hot carriers around the Γ and K points without momentum transfer. As we will show below, the different nature of the transient electronic excitation would directly influence the coupling to the phononic subsystem and lead to completely different scenarios in terms of structural dynamics.

In Fig. 2 we show the changes in the measured diffraction intensity recorded as a function of the delay time between the optical pump and the electron probe for the three high-symmetry directions, evidenced in the diffraction pattern in Fig. 1b, when using light at 800 nm (Fig. 2a-c) and at 400 nm (Fig. 2d-f). The transient change of the Bragg reflections unveils the non-equilibrium lattice dynamics. In the kinematic theory, the diffraction intensity of a given Bragg peak, $I_{[hkl]}$, is determined by the square modulus of the structure factor, $F_{[hkl]}$ (i.e., $I_{[hkl]} \propto |F_{[hkl]}|^2$), which depends on the atomic displacement with respect to the equilibrium positions:

$$F_{[hkl]} = \sum_j \xi_j \exp\left[i\boldsymbol{s}_{[\boldsymbol{hkl}]} \cdot \left(\boldsymbol{R_j} + \boldsymbol{u_j}(t)\right)\right] \tag{1}$$

Here, $\boldsymbol{u_j}$ defines the displacement of the j-th atom within the unit cell with respect to the equilibrium position $\boldsymbol{R_j}$, and $\xi_j$ is the atomic scattering factor. Local atomic rearrangements, as well as atomic vibrations, would induce a net nonuniform displacement within the lattice, resulting in a modification of the structure factor, and thus a transient variation of the diffraction intensity. A positive or negative change of the structure factor for the given Bragg reflection would thus cause an increase or a decrease of the diffraction intensity corresponding to specific atomic displacement triggered by the excitation of specific vibrational modes. Moreover, it is worth noting that what dominates the structure factor is the projection of the atomic displacement, $\boldsymbol{u_j}(t)$, on a particular crystallographic direction – defined by the scattering vector $\boldsymbol{s}_{[\boldsymbol{hkl}]}$ – via the scalar product $\boldsymbol{s}_{[\boldsymbol{hkl}]} \cdot \boldsymbol{u_j}(t)$. In this sense, modulation of the diffraction intensity also depends on the relative orientation of the atomic motion with respect to the scattering direction.

In the case of optical excitation at 400 nm (3.1 eV), the observed transient intensity change exhibits a progressive drop for all measured diffraction peaks with a time constant on the order of a few tens of picoseconds. This behavior is characteristic of semiconducting systems following the above band-gap excitation, where a large population of hot carriers is produced. The hot carriers cool down via direct coupling to hot optical and acoustic phonons, which in turn decay toward long-wavelength acoustic phonons. The increased dynamic disorder induced by the large-amplitude low-energy acoustic vibrations will consequently create a non-negligible loss of electron interference responsible for the usually oberseved decrease of diffraction intensity [19,20,22,24,25,35].

In the case of optical excitation at 800 nm (1.55 eV), the observed behavior is very different. Here, while the $\{1\bar{1}00\}$ Bragg peak still exhibits an intensity drop, the peaks associated with the $\{01\bar{1}0\}$ and $\{10\bar{1}0\}$ reflections show, instead, a sharp intensity increase on a time scale of a few picoseconds (4 – 5 ps) before recovering toward a negative change on a longer time scale of tens of picoseconds. Such early-time behavior indicates a positive increase of the structure factor. A similar increase of $F_{[hkl]}$ has been observed in the case of displacive lattice distortions induced in the excited state via the stabilization of a subset of phonons excited within the lattice. This has been observed – for instance – in the case of infrared resonant excitation of selective vibrational modes in strongly-correlated electron material YBCO [36], and anharmonic lattice distortions of model thermoelectric material SnSe [37]. Such a picture would be also consistent with the nature of the electronic excitation at 1.55 eV in $MoS_2$, where indirect $\Gamma_V$-$K_C$ and $\Gamma_V$-$Q_C$ excitons could be stabilized via the excitation of a subset of phonons with a well-defined momentum.

To quantitatively elucidate the transient atomic behavior of the lattice following the electronic excitation, we have calculated via density functional theory (DFT) the atomic details of the 18 normal vibrational modes allowed for a 2H-$MoS_2$ bulk crystal (further details on the DFT calculations are given in the Methods section) and determined the electron-phonon coupling matrix element for each mode at different electron and phonon momentum states across the Brillouin zone.

Fig. 3a-c report the electron-phonon matrix element calculated for each vibrational mode. For our analysis we have considered phonon modes at the Γ point (Fig. 3a), at the Q point (Fig. 3b), and at the K point (Fig. 3c) of the Brillouin zone. We have calculated the phonon excitation probability at each high-symmetry point (Γ, Q and K) associated with electronic transitions from the valence band states in Γ to the corresponding final states in the conduction band (at Γ, Q and K, respectively). In the calculations, momentum conservation for the indirect transitions is fulfilled by considering only phonons with the corresponding required momentum.

From the calculations in Fig 3a we immediately notice that in the case of direct excitation with following coupling to Γ phonons, the dominant contribution is given by the out-of-plane $A_{1g}$ phonon. This is in strong agreement with previous works that reported strong coupling of the A and B direct excitons at ~1.85 eV and ~2.1 eV, respectively, with $A_{1g}(\Gamma)$ out-of-plane phonon modes [5,6].

Different is the situation in the case of indirect transitions accompanied by the excitation of phonon modes at the K and Q points, shown in Fig. 3b-c. Here, besides the excitation of the out-of-plane $A_1'$ mode (equivalent to $A_{1g}$ in $D_{6h}$ symmetry), the main contribution to the lattice dynamics is given by in-plane (optical and acoustic) phonon modes with $E'$ symmetry (equivalent to $E_{2g}$ and $E_{1u}$ in $D_{6h}$ symmetry). This is also consistent with previous works [5,6] reporting that the favored coupling occurs with $E_{2g}^1$ modes when the exciton wave function has contributions from different parts of the momentum space (and not only Γ), such as in the case of high-energy C excitons at 2.7 eV.

The $E'$ modes correspond to in-plane motions where the Mo layer moves in opposite direction with respect to the two S layers. At low excitation (i.e. excited carrier densities below $1x10^{13}$ $cm^{-2}$), the system remains in the harmonic approximation and the atoms oscillates around their initial equilibrium position. This only leads to a non-zero mean square displacement of the atoms, which results in a negative change of the structure factor (see Fig. 4a and Supplementary section S1 for details).

In our case, we estimate an excited current density of about 7 x $10^{13}$ $cm^{-2}$ or above (see Supplementary section S2), and therefore we cannot neglect the effect of phonon anharmonicities of the excited $E'$ and $A_1'$ modes. As a result, the potential energy surface of the ground state becomes more flat and asymmetrical leading to a displaced equilibrium position following the photoexcitation (see Fig. 4b) [37, 38, 39]. Based on these considerations, we have then estimated for the different Bragg reflections the structure factor changes resulting from in-plane and out-of-plane displacements of S and Mo atoms following the direction of motion consistent with $A_1'$ and $E'$ modes (see Fig. 4c-e and Supplementary section S1).

Fig. 4c shows that the structure factor change calculated for out-of-plane displacements is identically zero for the $\{10\bar{1}0\}$, $\{01\bar{1}0\}$, $\{1\bar{1}00\}$ reflections. This directly derives from Eq. (1): the scalar product in the equation is zero since the scattering vectors of those reflections lie in the plane of the flake. Therefore, when $A_1'$ (and equivalently $A_{1g}$) is the dominant contribution to the electron-phonon coupling, such as in the case of direct transitions involving Γ phonons (see Fig. 3a), the only effect on the structure factor is a negative change resulting from the increased mean-square displacement of the atoms associated with the phononic excitation (see Fig. 4a). This picture is consistent with the experimentally measured decrease of the diffraction intensity for illumination at 400 nm, which results in the vertical excitation of hot carriers with zero transferred momentum, and thus the electron-phonon coupling is dominated by phonons at the Γ point. The incoherent deexcitation of such Γ-phonons will eventually involve also long-wavelength phonons at lower energies, leading to an increased mean-square atomic displacement and larger dynamical disorder [19,20,22,24,25,35].

Different is the case for in-plane atomic displacements. In Fig. 4d and 4e we show the structure factor changes calculated for in-plane displacements of S and Mo atoms (as indicated in the insets) for the $\{10\bar{1}0\}$, $\{01\bar{1}0\}$, $\{1\bar{1}00\}$ reflections. In the Supporting section S1 we also show the results obtained for a large range of possible S and Mo in-plane displacements along different directions of the unit cell (namely the two axis of the losanga forming the base of the unit cell). Such detailed analysis showed that when the deformation creates a distorted unit cell yielding a twisted ochtaedral coordination around the Mo atoms (as sustained by the excitation and anharmonicity of $E'$ modes), the structure factor change is positive for the $\{10\bar{1}0\}$ and $\{01\bar{1}0\}$ reflections, whereas it is negative for the $\{1\bar{1}00\}$ reflection.

By comparing the measured diffraction intensity changes for 800-nm illumination in Fig. 2 with the calculated relative structure factor changes for in-plane Mo and S displacements as shown in Fig. 4d-e, we can conclude that the transient diffraction intensity increase observed for Bragg reflections $\{10\bar{1}0\}$ and $\{01\bar{1}0\}$ is caused by such in-plane unit cell distortion as driven by the excitation and anharmonic effects associated with $E'$ phonon modes.

The combination of UED results, ab initio calculations, and structure factor calculations in the anharmonic regime allowed us to identify the microscopic nature and symmetry of the involved lattice vibrations coupling to the indirect excitons excited by infrared light at 1.55 eV. The nearly-resonant excitation of indirect $\Gamma_V$-$K_C$ and $\Gamma_V$-$Q_C$ excitons thus induces the excitation of $E'$ in-plane phonon modes with a well-defined momentum q = K, Q, necessary to conserve the momentum during a non-vertical (indirect) electron transition from the valence states to the conduction states. In this sense such phonon modes "stabilize" the formation of such indirect excitons. In other words, we have been able to directly determine that below-gap light absorption is possible via the excitation of K- and Q-phonon modes with $E'$ symmetry.

Our picture is further confirmed by the calculated Eliashberg function in bilayer $MoS_2$ [40], where the authors evidenced that, for the K point of the Brillouin zone, the phonon modes with the strongest electron-phonon coupling are both optical and acoustic with $E_{2g}$ symmetry. Our results are also consistent with recent investigations of electron-phonon interaction in $MoS_2$ via ultrafast visible/IR spectroscopy, where the authors observed the appearance of a well-defined feature in the frequency range 380-390 $cm^{-1}$ following excitation at 800 nm [41], which they attributed to the excitation of in-plane phonon modes with $E_{1u}$ symmetry.

The excitation of $E'$ phonons and their anharmonicity following photoexcitation initially drives the non-equilibrium distortion of the unit cells at early times. On the other hand, the long-term intensity decrease of the investigated Bragg reflections (see Fig. 2) – evolving on a time scale of tens of picoseconds – highlights the subsequent role of long-wavelength atomic motions in the thermalization process that causes the excitation of a large amount of low-energy acoustic modes.

Such a process induces a progressive increase of the mean-square displacements (see Fig. 4a), that leads to a negative change of the structure factor for all monitored Bragg reflections on a time scale of several tens of picoseconds, as observed experimentally (see Fig. 2) [35].

The $E'$ phonon excitation observed here in nano-flakes of $MoS_2$ for illumination at 800 nm is directly associated with the specific indirect-exciton nature of the wavelength-dependent electronic transition triggered in the system. Therefore, given the indirect character of such electronic states, we expect that the coherent structural response of the lattice would persist over long time scales. The preservation of such structural coherence can be directly seen by monitoring the transient change of the interatomic distance, shown in Fig. 4 for the case of the $[0\bar{1}10]$ direction (similar oscillatory behaviour of the scattering vector change is also present for the other Bragg reflections). We observed the presence of in-plane breathing oscillations with a frequency of 72.9 GHz – corresponding to a period of 13.7 ps. This is consistent with experimentally reported in-plane strain waves in multilayer $MoS_2$ observed at a frequency of about 50 GHz [42,43]. Moreover, considering that the longitudinal in-plane speed of sound in $MoS_2$ is $v$ = 7000 m/s [44] and the thickness of the investigated nanoflake is $t$ = 50 nm, this means a breathing oscillations period of $2t/v$ = 14.3 ps. This agrees well with the observed oscillatory period of 13.7 ps and confirms the in-plane breathing nature of the observed oscillations of the scattering vector for the monitored Bragg reflections. This coherent lattice oscillation can be seen as the result of an anharmonic coupling of the excited high-energy $E'$ in-plane Q- and K-phonon modes to low-energy acoustic vibrations [45] that will quickly relax into low-frequency in-plane strain waves.

## CONCLUSIONS

In this paper we have tackled the key problem of the role played by the atomic degrees of freedom in the non-equilibrium dynamics of indirect excitons in $MoS_2$, the prototypical transition metal dichalcogenide. Previous theoretical and experimental studies had mainly focused on the exciton-phonon dynamics in high-energy direct excitons involving zone-center phonons. Here, we have investigated the many-body exciton physics of zone-edge states, which is crucial to fully exploit the valley-selective excitations in these materials, and unraveled their relevant ultrasmall and ultrafast spatiotemporal scales through a unique combination of ultrafast electron diffraction, *ab initio* calculations and structure factor calculations in the anharmonic regime. In particular, we have identified the excitation of in-plane Q- and K-phonon modes with $E'$ symmetry as an important contribution to nearly-resonant excitation of low-energy indirect excitons in $MoS_2$. Our UED results can be explained by considering the effect of anharmonicity of such $E'$ phonons, inducing an in-plane Mo and S displacements compatible with a unit cell distortion yielding a twisted ochtaedral coordination around the Mo atoms. Because the observed mechanisms have significant impact on the optical and electronic properties of these materials in non-equilibrium conditions, likely setting an intrinsic limit to carrier mobility and coherence loss, our results thus not only provides the direct observation of the microscopic nature and symmetry of the involved lattice changes, but would also contribute to further improving the material performance.

## ACKNOWLEDGEMENTS

J.H. and G.M.V would like to thank the late Prof. Ahmed H. Zewail for his support and guidance in the initial phases of this work and for providing access to the UED facility at Caltech. J.H. and G.M.V would like to thank Dr. Haihua Liu for EELS characterization of the $MoS_2$ samples. J.H. acknowledges support from the Science Challenge Project (No. TZ2018001). G.M.V. and B.F. acknowledge support from the SMART-electron project that has received funding from the European Union's Horizon 2020 Research and Innovation Program under Grant Agreement 964591.

## METHODS

### Preparation of $MoS_2$ nano-flakes

$MoS_2$ nano-flakes used here are mechanically exfoliated from a 100% natural $MoS_2$ crystal provided by SPI Supplies/Structure Probe, Inc.. In detail, we use scotch tape to exfoliate thin flakes off the bulk $MoS_2$ after numerous repetitions, and then immerse the tape with $MoS_2$ flakes in acetone. After 3 - 4 minutes, we shake the tape in the acetone to detach $MoS_2$ flakes and then fish them by using 200-mesh Cu TEM grids. Here, we visually pick up half-transparent or nearly transparent $MoS_2$ flakes as the sample. To remove the acetone, we rinse the TEM grid with MoS2 many times by using deionized water. Finally, the grids are dried in the air for UED measurements. Electron energy loss spectroscopy (EELS) has been used to estimate the thickness of the produced nanoflakes, yielding a value on the order of 50 nm.

### Ultrafast electron diffraction (UED)

The UED measurements have been performed at Caltech in the laboratory of the late Prof. Ahmed H. Zewail. The output of a Ti:Sapphire regenerative amplifier ($\lambda$ = 800 nm, 100 fs, 2 kHz) was separated in two portions: one represents the optical pump beam that is responsible for exciting the sample, which was used either as fundamental at 800 nm or to generate the second-harmonic ($\lambda$ = 400 nm) via a BBO crystal; the other was used to generate ultraviolet pulses by third-harmonic generation ($\lambda$ = 266 nm), which irradiate a $LaB_6$ photocathode to produce ultrafast electron pulses. The diffraction patterns generated by electron scattering from the samples were recorded in far-field on a gated CCD detector and monitored as a function of the delay time between pump and probe. For the investigation of $MoS_2$ nanoflakes, we adopted a normal incidence geometry (transmission scheme), tracking the transient behavior of the in-plane reflections. For both light beams (800-nm and 400-nm) we used a linear polarization. Both beams were p-polarized (i.e. the polarization direction is parallel to the scattering plane), and they both arrived on the sample surface at a 45 degrees angle with respect to the normal to the sample surface.

To minimize space charge effects, an electron pulse was designed to contain less than 300 electrons, giving a sub-ps pulse duration. Particular care was taken for the calibration of the zero-delay time, i.e. the time at which the electron and laser pulses simultaneously arrived on the sample. We used multiphoton ionization from a metallic surface, which created a transient and localized plasma synchronous with the laser excitation able to modify the electron pulse's position and spatial profile ("plasma lensing effect"). Finally, we note that within the adopted fluence range transient electric field effects on the measured diffraction pattern were negligible and did not affect the measured dynamics [46].

**DFT calculations of phonon modes and electron-phonon coupling matrix elements**

Band structure calculations are carried out with the Vienna Ab-initio Simulation Package (VASP) [47] based on the Density Functional Theory with projector-augmented-wave (PAW) [48,49] pseudopotentials. The generalized gradient approximation (GGA) with the Perdew-Burke-Ernzerhof (PBE) parametrization is used for the exchange and correlation functional [50]. The computations are based on a unit cell consisting of 6 atoms with an 8×8×8 k-point mesh within Monkhorst-Pack scheme and cutoff energy of 500 eV. The convergence criteria for total energies and forces are $10^{-8}$ eV and $10^{-6}$ eV/Å, respectively. The bulk geometry used to simulate $MoS_2$ has a relaxed lattice parameter of a = 3.18 Å, b = 3.18 Å, c = 13.13 Å, α = 90°, β = 90°, γ = 120°.

The phonon dispersion is calculated using the direct force-constant method implemented in the PHONOPY code [51]. In this method, a specific atom is displaced to calculate the induced Hellmann Feynman forces on itself and its surrounding atoms. By collecting the Hellmann Feynman forces one can construct the dynamical matrices that lead to the phonon properties and thermodynamic functions within the framework of lattice dynamics and harmonic approximation. Because of the effects of image atoms due to the periodic boundary conditions, larger supercell is needed to make sure all calculated phonon frequencies are well converged. We employed a 4×4×2 supercell with 192 atoms for phonon calculations which are well converged in this work.

The electron-phonon matrix elements have been calculated by using the density-functional perturbation theory (DFPT) [52] as implemented in the QUANTUM ESPRESSO code [53]. The parameters and convergence criteria have been set as in the VASP calculations and PAW sets have been also used.

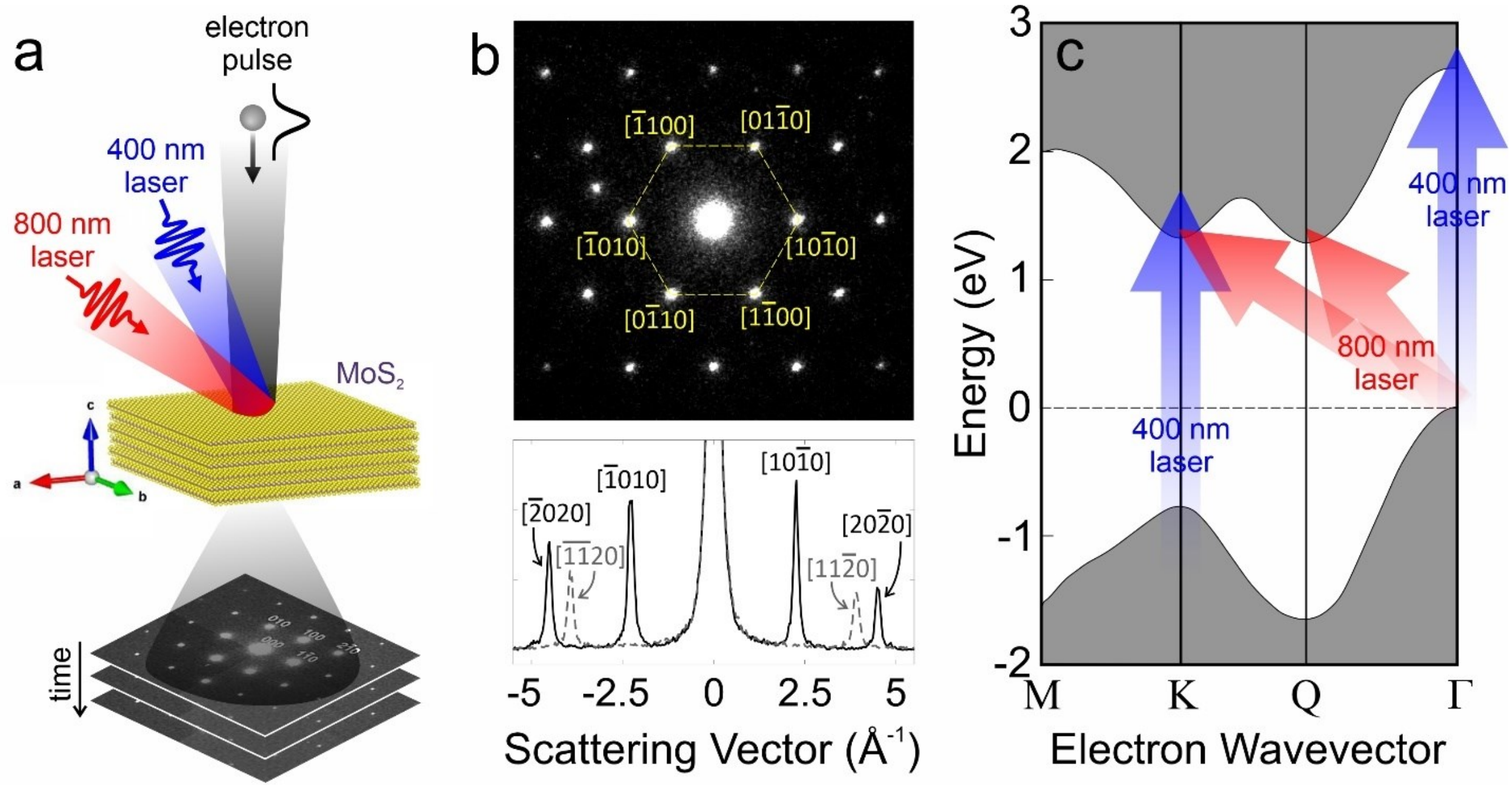

**Figure 1. UED experiment on nanoflakes of $MoS_2$. (a)** Schematic representation of the UED experiment where ultrashort electron pulses are focused in normal incidence on the $MoS_2$ flake and the dynamics is initiated by ultrafast laser pulses with a variable wavelength (either 800 nm or 400 nm). The diffracted electrons are then recorded in stroboscopic mode at different delay times between the excitation laser and the probing electron pulse. **(b)** Representative diffraction pattern measured before laser excitation, showing the in-plane hexagonal pattern typical of a $D_{6h}$ crystal where we can identify three high-symmetry lattice plane families: $\{1\bar{1}00\}$, $\{01\bar{1}0\}$, and $\{10\bar{1}0\}$. **(c)** Schematic representation of the electronic band structure for a 2H-$MoS_2$ bulk crystal showing an indirect band gap of 1.29 eV at the Q point along the Γ-K direction of the Brillouin zone; the direct gaps in Γ and in K are, instead, above 2 eV. The blue and red arrows represent allowed transitions following optical excitations at 400 nm (3.1 eV) and 800 nm (1.55 eV), respectively.

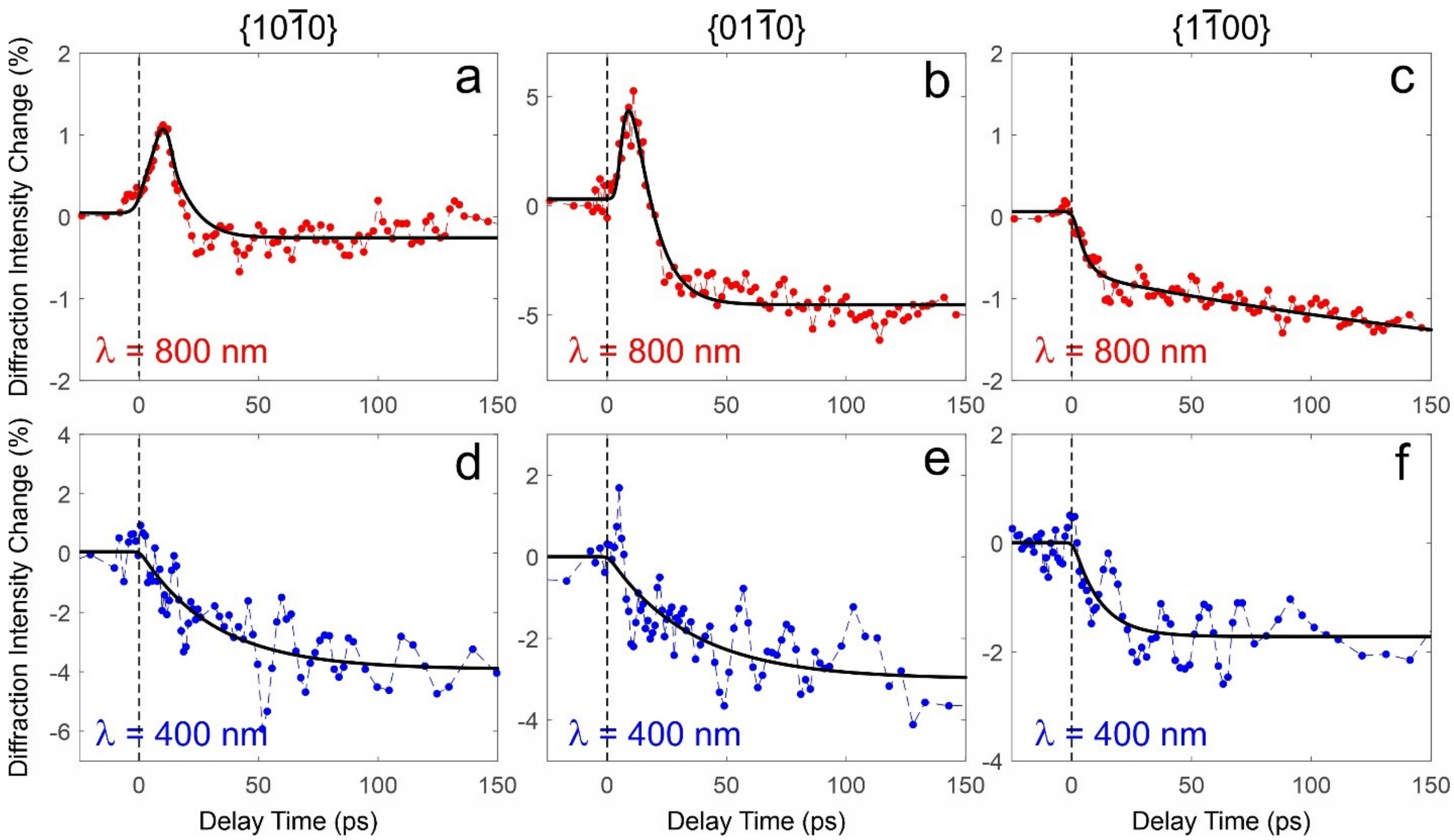

**Figure 2. Transient lattice evolution following the optical excitation.** The measured diffraction intensity changes are recorded as a function of the delay time between the optical pump and the electron probe for the three high-symmetry directions evidenced in the diffraction pattern in Figure 1b when using light at 800 nm and at 400 nm. **(a)** $\{10\bar{1}0\}$ reflection at 800 nm; **(b)** $\{01\bar{1}0\}$ reflection at 800 nm; **(c)** $\{1\bar{1}00\}$ reflection at 800 nm; **(d)** $\{10\bar{1}0\}$ reflection at 400 nm; **(e)** $\{01\bar{1}0\}$ reflection at 400 nm; **(f)** $\{1\bar{1}00\}$ reflection at 400 nm.

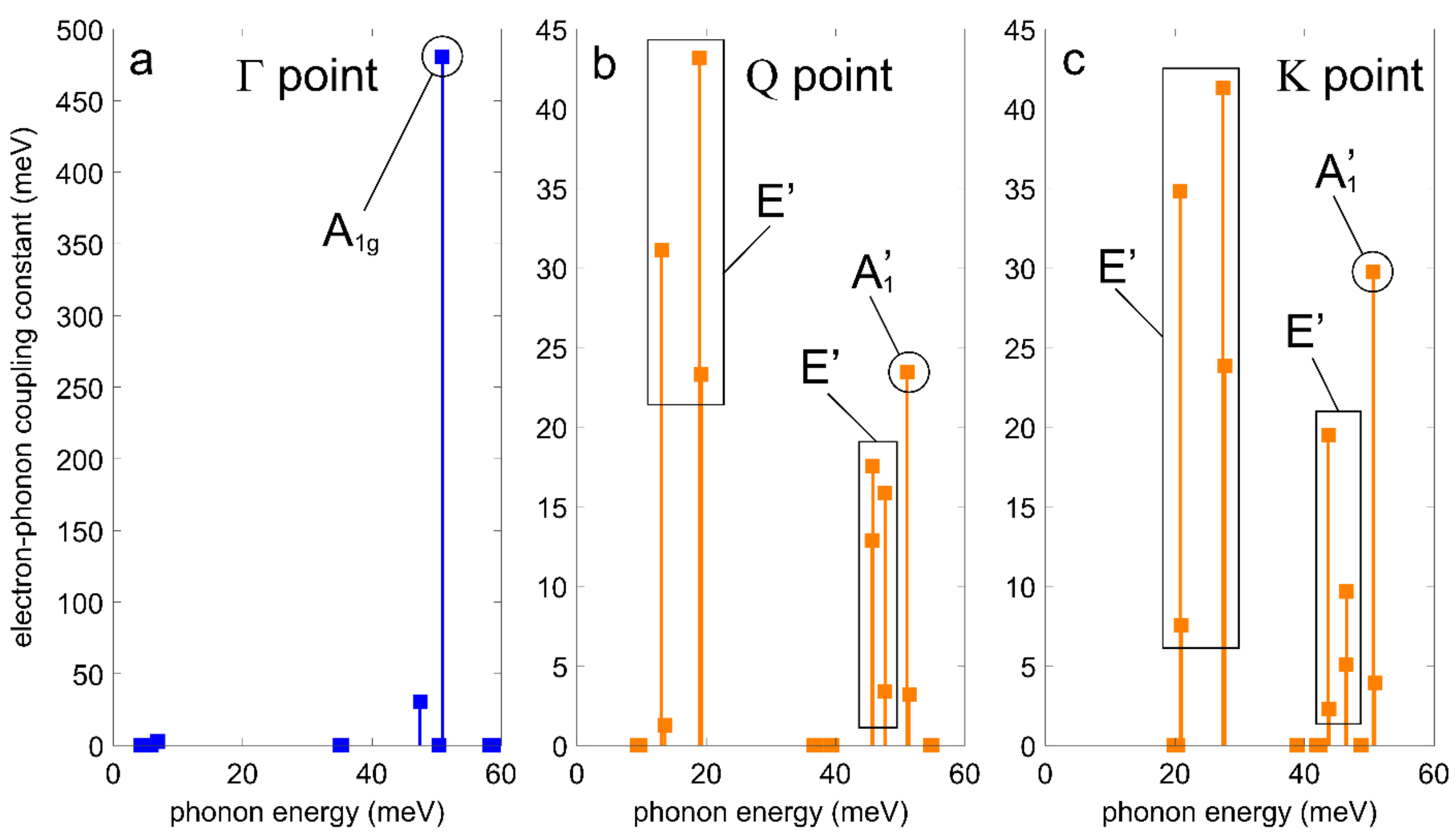

**Figure 3. Calculations of electron-phonon coupling matrix elements.** The electron-phonon coupling matrix elements are calculated via density-functional perturbation theory (DFPT) for each vibrational mode and specific momentum state across the Brillouin zone. We consider: direct electron transitions $\Gamma_V$-$\Gamma_C$ with coupling to Γ phonons (panel a); indirect $\Gamma_V$-$Q_C$ electron transitions with coupling to Q phonons (panel b); indirect $\Gamma_V$-$K_C$ electron transitions with coupling to K phonons (panel c).

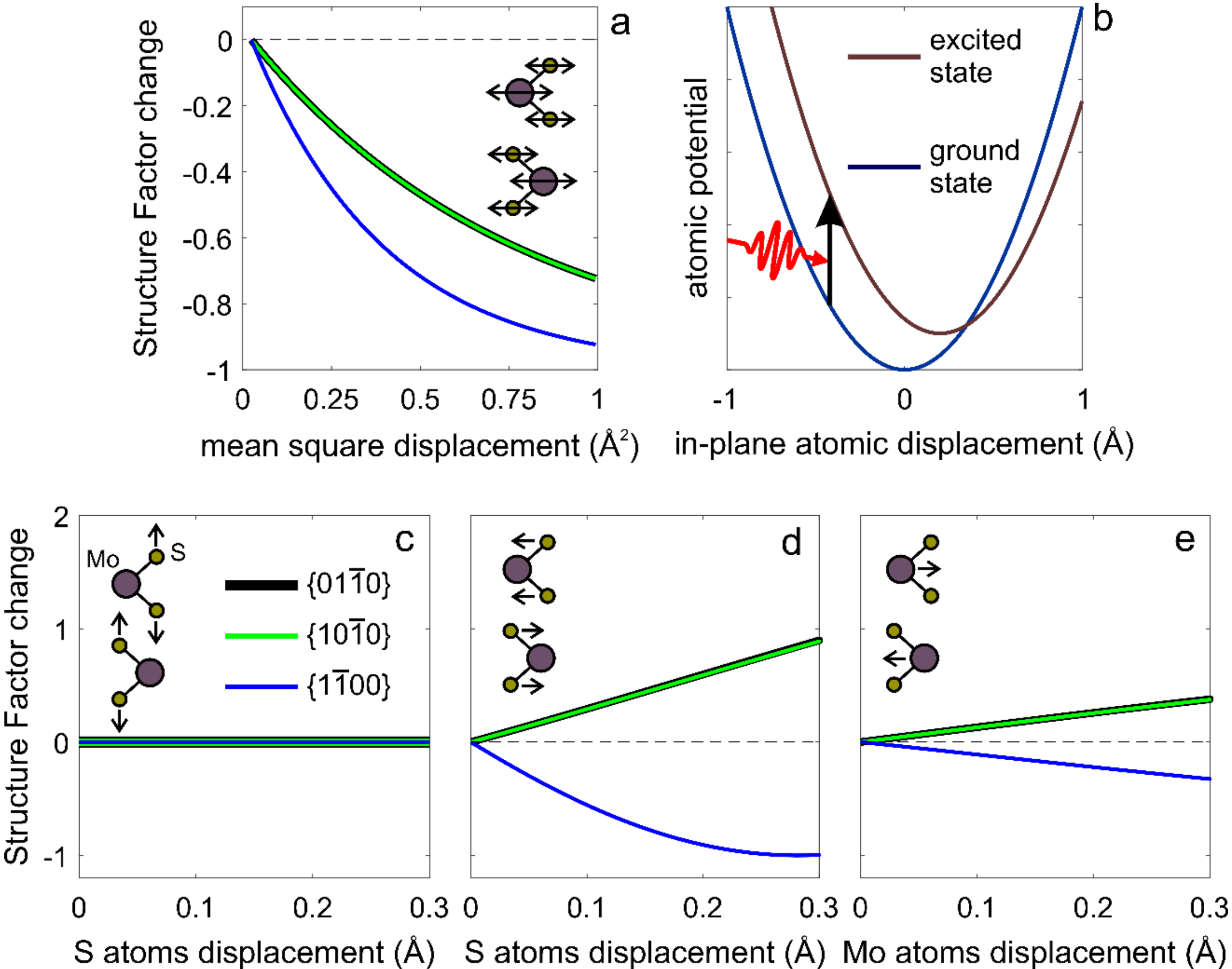

**Figure 4. Structure factor changes driven by harmonic and anharmonic lattice displacements.** (a) Calculated change of the square modulus of the structure factor, $|F_{[hkl]}|^2$, for an increasing mean square displacement of the atoms around their initial equilibrium position (harmonic regime). (b) Schematic representation of the ground state and excited state potential energy surfaces as modified by phonon anharmonicities of the excited modes leading to a displaced equilibrium position following the photoexcitation. (c) Estimated change of the square modulus of the structure factor, $|F_{[hkl]}|^2$, for the $\{10\bar{1}0\}$ (green curve), $\{01\bar{1}0\}$ (black curve), $\{1\bar{1}00\}$ (blue curve) Bragg reflections resulting from out-of-plane displacements of S atoms (panel c), in-plane displacements of S atoms (panel d) and in-plane displacements of Mo atoms (panel e) following the direction of motion consistent with $A_1'$ and $E'$ modes.

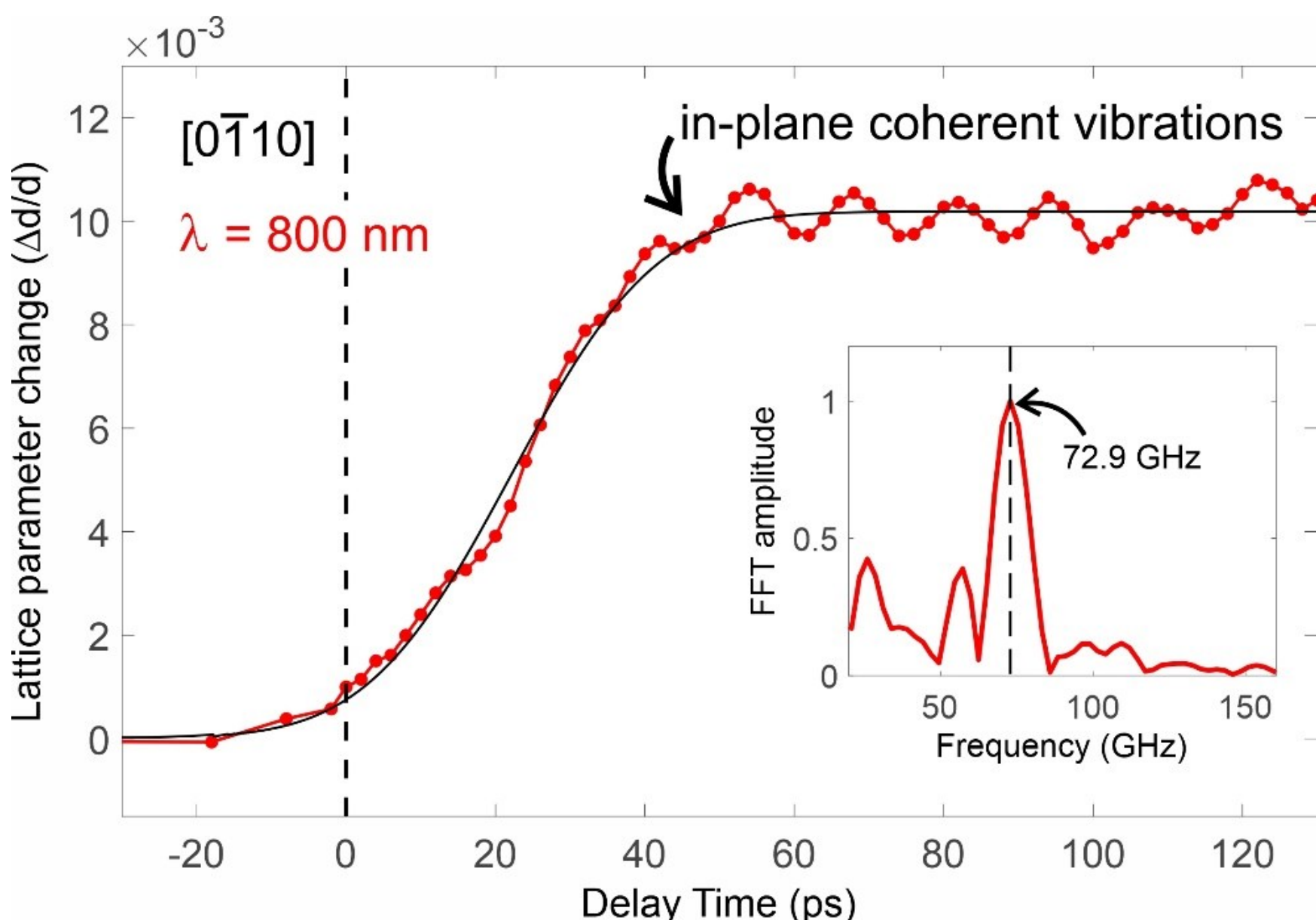

**Figure 5. Long-term coherent lattice dynamics.** Transient change of the interatomic distance measured for the case of the $[0\bar{1}10]$ direction, showing in-plane breathing oscillations with a frequency of 72.9 GHz (period of 13.7 ps). This behavior is consistent with a coherent structural response of the lattice persisting over long time scales.

[14] S. Lahme, C. Kealhofer, F. Krausz, and P. Baum, Femtosecond single-electron diffraction, *Struct. Dyn.* **1**, 034303 (2014).

[15] I-Cheng Tung, Aravind Krishnamoorthy, Sridhar Sadasivam, Hua Zhou, Qi Zhang, Kyle L. Seyler, Genevieve Clark, Ehren M. Mannebach, Clara Nyby, Friederike Ernst, Diling Zhu, James M. Glownia, Michael E. Kozina, Sanghoon Song, Silke Nelson, Hiroyuki Kumazoe, Fuyuki Shimojo, Rajiv K. Kalia, Priya Vashishta, Pierre Darancet, Tony F. Heinz, Aiichiro Nakano, Xiaodong Xu, Aaron M. Lindenberg & Haidan Wen, Anisotropic structural dynamics of monolayer crystals revealed by femtosecond surface x-ray scattering, *Nature Photonics* **13**, 425–430 (2019)

[16] Frigge, T. et al., Optically excited structural transition in atomic wires on surfaces at the quantum limit, *Nature* **544**, 207–211 (2017).

[17] S. Vogelgesang, G. Storeck, J. G. Horstmann, T. Diekmann, M. Sivis, S. Schramm, K. Rossnagel, S. Schäfer & C. Ropers, Phase ordering of charge density waves traced by ultrafast low-energy electron diffraction, *Nature Physics* **14**, 184–190 (2018)

[18] M. R. Otto, J.-H. Pöhls, L. P. René de Cotret, M. J. Stern, M. Sutton, and B. J. Siwick, Mechanisms of electron-phonon coupling unraveled in momentum and time: The case of soft phonons in $TiSe_2$, *Science Advances* **7** (2021)

[19] Ehren M. Mannebach, Renkai Li, Karel-Alexander Duerloo, Clara Nyby, Peter Zalden, Theodore Vecchione, Friederike Ernst, Alexander Hume Reid, Tyler Chase, Xiaozhe Shen, Stephen Weathersby, Carsten Hast, Robert Hettel, Ryan Coffee, Nick Hartmann, Alan R. Fry, Yifei Yu, Linyou Cao, Tony F. Heinz, Evan J. Reed, Hermann A. Dürr, Xijie Wang, and Aaron M. Lindenberg, Dynamic Structural Response and Deformations of Monolayer $MoS_2$ Visualized by Femtosecond Electron Diffraction, *Nano Lett.* **15**, 6889−6895 (2015)

[20] Ming-Fu Lin, Vidya Kochat, Aravind Krishnamoorthy, Lindsay Bassman, Clemens Weninger, Qiang Zheng, Xiang Zhang, Amey Apte, Chandra Sekhar Tiwary, Xiaozhe Shen, Renkai Li, Rajiv Kalia, Pulickel Ajayan, Aiichiro Nakano, Priya Vashishta, Fuyuki Shimojo, Xijie Wang, David M. Fritz & Uwe Bergmann, Ultrafast non-radiative dynamics of atomically thin $MoSe_2$, *Nature Communications* **8**, 1745 (2017).

[21] Linlin Wei, Shuaishuai Sun, Cong Guo, Zhongwen Li, Kai Sun, Yu Liu, Wenjian Lu, Yuping Sun, Huanfang Tian, Huaixin Yang, and Jianqi Li, Dynamic diffraction effects and coherent breathing oscillations in ultrafast electron diffraction in layered 1 T-TaSeTe, *Structural Dynamics* **4**, 044012 (2017)

[22] J Hu, GM Vanacore, A Cepellotti, N Marzari, AH Zewail, Rippling ultrafast dynamics of suspended 2D monolayers, graphene, *Proceedings of the National Academy of Sciences* **113**, E6555-E6561 (2016).

[23] J Hu, GM Vanacore, Z Yang, X Miao, AH Zewail, Transient Structures and Possible Limits of Data Recording in Phase-Change Materials, *ACS Nano* **9**, 6728-6737 (2015)

[24] GM Vanacore, J Hu, W Liang, S Bietti, S Sanguinetti, AH Zewail, Ultrafast atomic-scale visualization of acoustic phonons generated by optically excited quantum dots, *Nano Letters* **14**, 6148-6154 (2014)

[25] GM Vanacore, RM Van Der Veen, AH Zewail, Origin of Axial and Radial Expansions in Carbon Nanotubes Revealed by Ultrafast Diffraction and Spectroscopy, *ACS Nano* **9**, 1721-1729 (2015)

[26] W Liang, GM Vanacore, AH Zewail, Observing (non)linear lattice dynamics in graphite by ultrafast Kikuchi diffraction, *Proceedings of the National Academy of Sciences* **111**, 5491-5496 (2014)